# Some Information Is too Dangerous to Be on the Internet*

Vincent C. Müller
Princeton University, Hellenic Studies &
Anatolia College/ACT
www.typos.de

24[th] February 2006

This paper investigates a problem about freedom of information. Although freedom of information is generally considered desirable, there are a number of areas where there is substantial agreement that freedom of information should be limited. After a certain ordering of the landscape, I argue that we need to add the category of 'dangerous' information and that this category has gained a new quality in the context of current information technology, specifically the Internet. This category includes information the use of which would be morally wrong as well as some of what may be called 'corrupting' information. Some such information should not be spread at all and some should be very limited in its spread.

*Keywords:* Censorship, corruption, dangerous information, illegal information, freedom of information, intellectual property, spread of information, secrets.

---

## 1. Limits to the Freedom of Information

### 1.1. Spread and Use of Information

Though freedom of information is an important value, there are a number of well established ethical arguments that warrant limits on the freedom of information. I propose that they can broadly be characterised to concern the *spread* or the *use* of information.[1]

In our context, we can assume information to cover only *content* that could become knowledge of a person, that is it is either 'information *that*' (something is the case) or 'information *how*' (to do something). Note that when talking about 'information that/how', I do not mean to limit information to what can, in principle, be expressed linguistically: Information can also concern *that* something looks like this, tastes like this, sounds like this, has this magnetic field, etc.

It is important to realise that only a small part of our speech and our use of information technology concerns the spread or use of information in this sense. When we perform speech acts such as, for example, asking someone for advice, insulting someone, giving orders, or saying "I do" at a wedding ceremony, we want to *do* something, something other than spread information (though, of course, in all these cases information is actually spread *also*). Accordingly, there are any number of speech acts that can be done via information technology but that are neither spread nor use of information. What is more, there are many kinds of further acts that can be done via information technology now (and who knows what else we will be able to do in the future): to transfer money, to vote, to buy something, to sign a contract, to rob a bank, to fire a missile, etc. etc. What is ethically relevant in these acts is not the spread or use of information but what is done, and what could be done by quite different means, so our ethical arguments here shall not deal with these kinds of issues - and perhaps information ethics quite generally should not concern itself with these.[2]

---

[1] The general and terminological clarifications of these first sections may well be reinventing the wheel; indeed I would hope that they are, only that I have not found the previous invention. The legal issues are well defined and a good specialised dictionary provides the necessary clarifications, e.g. Creifelds 1986 (see also Council of Europe 2001). A good start is Severson (1997, 134): "The Principles of Information Ethics. 1. Respect for intellectual property. 2. Respect for privacy. 3. Fair representation. 4. Nonmaleficience (or 'doing no harm')." DeGeorge (2003a) offers a survey of problems business ethics that relate to information technology.

Now, limiting ourselves to the spread or use of information, there are several kinds of cases where this action is traditionally considered problematic – and thus will continue to be problematic when performed via information technology. In fact, the issue of spreading information has been seriously aggravated by modern Information Technology where we now do not have to transport and duplicate atoms, but bits. This allows duplicates that are identical, unlimited in number, transported at high speed, that are cheap and easy to make - as well as hard to identify and to trace. We will see below how this creates a new quality of danger for some kinds of information.

### 1.2. Spread

In the case of *spreading information*, the mere fact that a person acquires information is considered problematic, even if they make no *use* of it. 'Acquisition' of information is meant to be a stronger notion than 'access' here: access requires merely that a person *could* acquire information if he/she so desired, acquisition is the active understanding of that information. The result of acquisition I simply call 'having' information - whether or not it goes with belief in the truth of that information. So, if I steal your diary, I gain access to the information, but I only acquire it once I open and read the diary (unless upon opening it, I discover it is written in an indecipherable script, in which case I cannot acquire the information and never had access to it). I will limit the discussion to acquisition, even though in many cases even mere access is considered problematic - but these are a subset of the cases where acquisition is problematic.

The spread of information does not require an active effort to disseminate the information; it can also be done negligently. What matters here is acquisition, not the intention to disseminate. So, which are the kinds of information that should not be spread?

(1) The first kind of such information can be called 'secrets'. These fall into two kinds, (a) those secrets to which a particular person has a right, known as

---

[2]    Accordingly, I shall not deal with what quite ordinary criminals are now doing with information technology. For illustration: 'http://forum.cvv.ru This is where all the great criminal minds come together.' (From a 'spam' e-mail to me, 04 Nov 2003, proposing

'privacy': certain individuals (or groups) are thought to have particular rights concerning information *about these same individuals* (or groups). The right to privacy is legally protected (in standard legal systems) as part of 'personality rights'. A large part the discussions in computer ethics concern this right, which is particularly threatened by technical developments. (b) Less frequently discussed is the issue of what may be called 'non-private secrets', of information that natural and legal persons (or institutions) may have, but do not wish to spread. These secrets are not personal and are not owned, so several people may wish to keep the same secret. It is commonly thought to be a right of a person to keep a secret, and thus wrong to reveal a secret against the will of that person. If the secret has marketable value or effects, in particular, it gains special legal protection as a 'trade secret'. Virtually all organisations have such secrets, some operate sophisticated classification and security systems (esp. states, the military, etc.). Secrets are also the basis for multi-user operating systems that restrict certain types of file-access to certain people.

(2) The second kind of information that is thought to require restriction is offensive or defamatory information about a person (or group). Here, information spread should be restricted if the 'honour' of the persons in question is not respected, i.e. if the information is degrading and not verifiably true (it is then known as slander or libel).[3] This information may also concern deceased people, groups of people or symbols (such as the flag of a nation or a religious symbol like the crucifix).

(3) A third kind are lies, that is, information that is meant to deceive, normally a known falsehood. Though lying is generally considered to be ethically problematic, it is normally not sufficiently problematic to warrant legal restrictions - indeed there are cases where lying is considered ethically permissible or even demanded. However, there are cases where it is illegal, e.g. when it intends to damage someone else's financial state (then called 'fraud') or when it has other

---

all sorts of profitable criminal activities with the help of computing technology. Site now defunct.)

[3] I ignore the problem of what constitutes an 'insult', which may be illegal even if the information is true, provided it is meant to degrade the person concerned. (As when calling a someone a 'thief' or a 'pig'.) I think, in these cases the problem does not lie with the information, the content, but with the intention in which it is used and the effect it has on the person concerned. (Arguably, insults should be part of the category in question here.)

intentions considered sufficiently problematic (e.g. in Germany it is illegal to claim that the Holocaust never happened, so called "Auschwitz-lie").

(4) Then, there is 'corrupting information' that may not be spread to particular people, such as minors, because of the presumed detrimental effect on their character. We shall come back to this problem in the next section.

### 1.3. Use and Personal Rights

(5) A different category is information which can be *spread* without ethical problems, but not *used* by anyone without special rights or permissions. Restrictions on the *use* of information always concerns information that is owned (e.g. patents, copyrights, trademarks) by natural or legal persons who thus obtain the right to restrict its use by others, or at least its use for a profit.[4] These rights on the use of information plus the rights on trade secrets (2b) are jointly known as 'intellectual property rights' (cf. Spinello 2003).

All five issues above are considered to be due specific 'rights' and are well entrenched in standard codified legal systems.[5] In such legal systems, these rights are subject to restrictions and can be overruled under particular circumstances: e.g. the right to privacy for the purpose of law enforcement. The discussion of these details is of considerable interest to experts in ethics and law.

Note that the spread and use of information is limited by the rights of *particular* persons. These persons are A) the *subject matter* of this information (e.g. those who have rights to privacy [1a] or to personal respect [2, 3]) or B) the *rightful owner* of a particular piece of information (a non-personal secret [1b], or a patent etc. [5]). If these persons who are the subjects or the owners of the information give their consent, nothing is thought to be problematic with spreading or using that information. These persons have special rights to this information and they can decide to spread it.

In the special case of information that is 'corrupting' for minors [4], the state acts on their behalf, protecting what it takes to be perceived best interest and thus restricts information spread to them accordingly. Also, in some cases of lies [3],

---

[4] In some cases, the spread is *at the same time* also a use of information, such as in the spread of copyrighted material (the copying of a text, etc.). Clearly, however, it is the aspect of use that is considered problematic here, not the spread. (On copyright, see the US 'Digital Millenium Copyright Act'.)

[5] The Council of Europe (2001) convention has a useful survey of the main categories of cybercrime that is violating the rights mentioned here.

the state acts on behalf of persons or groups whose rights are thought to be infringed.

In all these cases, there are fruitful discussions about the extent to which the spread or use of information should be restricted (sometimes not so fruitful, as when the same people demand total freedom of information from others and protection of privacy for themselves). In all these cases, the problems have been exacerbated and altered by the advent of information technology.

### 1.4. What Should not Be Used Should not Be Known

I want to propose that there is a further category of information, the spread and use of which should be restricted. This category is different from the above in that it is not the result of the rights of particular individuals that stand in a particular relation (of ownership or of being the subject matter) to the information. Rather, I propose that a) there is information, the use of which is wrong and b) there is information which is such that its spread is wrong and c) that the spread of such information is wrong *because* its use is always wrong: so the classes a) and b) are co-extensional. If there is information the spread or use of which is wrong even though there are no persons with particular rights on this information (if a] or b] is true), then the spread or use of this information is wrong for anybody, not just for particular persons, as in the standard cases discussed above.

Is it correct to say that the spread of such information is wrong because its use is always wrong? Let us assume that we found a piece of information the use of which is always wrong. It does not seem hard to show in principle, that there could be information which should not ever be used: Imagine, you found a way to kill all life on Earth at a stroke. I would contend that using this information would be wrong, whatever you are trying to achieve.[6] So, we want to avoid use of the information. If there is no further reason why spread of the information is desirable, then it would follow that the spread of that information is wrong, too - as c) above claimed. I do not see any such further reason, apart from a general principle that the spread of information should be free - and that is precisely the principle at issue here. [So, I provide an argument only if you want to grant that there could be reasons to doubt that principle. The traditional reasons in the previous section provide some argument in this direction. If you hold freedom of information above other values that might be violated by the use of the information in question, I have no argument to offer, at this point.][7]

---

6  This could be contentious only if you think that there can be something that is valued higher than life on Earth, which would have to be something beyond physical existence on Earth (e.g. on another planet or in some metaphysical realm, like the Christian 'heaven'.)

## 2. What Should not Be Known Should not Be Spread

Concerning the problem of spreading information, candidates include: A) information known to be false, B) 'corrupting' information, C) dangerous information. A) The spreading of information that is known to be false (normally) aims at deception and is dishonest. This seems to warrant a characterisation as 'immoral' but it is not commonly assumed to be sufficient for motivating legal restrictions (i. e. the use of state force to prevent such spread). The freedom of expression typically covers the expression of known falsehoods - but note that personality rights sets some limits to this freedom, as mentioned above under 'libel'.

B) Many people believe that certain kinds of information have the property of having a negative effect on the character of persons who become aware of this information. This 'corrupting' effect is particularly attributed to pornographic material and to the depiction of cruelty, violence, etc. It is debatable whether such effects really occur and whether such effects warrant legal limitations on the spread of this information. Different societies have responded differently to these challenges - though it has to be said that the vast majority of states currently have some sort of legislation limiting the spread of certain such information, at least to certain people (minors).

The types A) and B) are the ones commonly discussed in the context of 'censorship' - usually with those advocating some forms of censorship in the defence (cf. 'Index on Censorship' and Caslon 2004). Is is type C) that, in my opinion, has not been given the attention it deserves.

### 2.1. Censorship and Corruption of Character

It is easy to ridicule and then dismiss the notion that information can be corrupting a person's character. There are so many examples in history where moral apostles have warned against the corrupting effect, especially on the youth, of some text or image - and now we would look at these moral apostles with an understanding smile, at best. This happens mostly because the effect in question is not considered (sufficiently) problematic any more, as in homosexuality, masturbation, weakening of religious faith, etc.

---

[7] There are many sites on the Internet that violate the principles defended here. Of course, I cannot indicate them here, for pains of contradicting myself.

The basic idea, however, that information can have an effect on a person's personality is hard to deny and is at the heart of much pedagogical advice and many legal restrictions, even today. The question is which effects one considers bad, corrupting, and how serious their influence is supposed to be - or, to put it the other way 'round, how little trust in the character of the persons concerns one has. We still restrict the access of minors to explicitly sexual and to violent[8] material, in fact, in many places of the world, these restrictions do not just apply to minors (the US restricts 'obscene material')[9]. Furthermore, there is still material that is thought to be sufficiently corrupting to warrant outlawing its spread, even to adults. This applies to particular forms of pornography (paedophilia, necrophilia, zoophilia), which are illegal pretty much world-wide - the motivation here seems to be partially a concern for the real victims in the production of (some of) the material and partially the effect on the consumers of the information, which is, in turn deemed dangerous (encouraging paedophile tendencies might lead to more child abuse). Such restrictions also apply to some forms of political propa-

---

[8] Some sites specialise on the publication of such material, trying to be as offensive as possible, while apparently remaining just about within the limits of US law. They claim to be extremely popular: "We get about 15 million hits a day, this totals 250GB of downloads during that period. About 250,000 unique individuals will visit the site during the course of a day. It's quite insane. According to the PCDATA web rating figures, Rotten is more popular than the New York Times website." (Online April 2005. Find the source if you like, but I do not recommend it.)

[9] This provision is not enforced, recently. 'Obscene' was defined in the United States Supreme Court 1973 decision Miller v. California as follows: 'Before sexual material can be judged obscene and therefore unprotected by the First Amendment, a judge or jury must determine: 1. that the average person, applying contemporary community standards, would find that the work, taken as a whole, appeals to prurient interest; 2. that the work depicts or describes, in a patently offensive way, sexual conduct specifically defined by the applicable law; and 3. that the work, taken as a whole, lacks serious literary, artistic, political, and scientific value.'

In the UK, activities around pornography are observed and regulated by the quasi-governmental organisation 'Internet Watch Foundation', www.iwf.org.uk

ganda that particular states consider especially dangerous, e.g. (for Europe) Nazi propaganda in Germany, racist propaganda in the UK and France[10].

Obviously, different societies and different people will disagree on which information is sufficiently corrupting to warrant restrictions on their distribution and access (due to disagreements on what is dangerous and on how much should be regulated). Restrictions on the grounds of 'corruption' can properly called 'censorship'.[11] Such censorship may have the best possible intentions, but it remains a violation of people's autonomy of information. I do not see a sufficient reason to practice censorship as long as we cannot be *certain* that some particular information will be harmful. (Even the case of minors is doubtful, though there we can presumably argue that they need to be protected precisely because they are not yet fully autonomous persons.) Note that we can be certain that an information is harmful if it violates one of the personality rights mentioned at the outset.

Accordingly, no form of political censorship is covered by the argumentation presented here. Al Qaeda should have the right to a web-site to defend their views. It is only if they spread bomb-making guidelines or the like that this would fall under the category of 'dangerous information' proposed here.

### 2.2. A Note on Technicalities and Links

The problems of enforcing any restriction of spread on the Internet are, of course, formidable. National responsibilities currently still exist for any site on the Internet (though there are attempts to circumvent this via ships on the open sea or other places with unclear sovereignty), but how can this be enforced? Anscombe (2004) proposes a national black list for the UK, controlled by an independent collecting organisation - perhaps his own? But should law enforcement not remain a state prerogative? Though many states (and the EU) have made steps in this direction, even the cases where there is significant agreement on the legal side (e.g. child pornography), the practical issues are largely unresolved.

---

[10] France: Forum des droits sur l'internet (2004). In many of these cases, not only the spread but also the possession of the material is illegal. (This problem goes beyond the confines of this paper.).

[11] Note that there is also a traditional use of this term which is narrower, covering only the requirement for state approval *prior* to publication, as when a state requires permissions to print a book.

One aspect of restrictions would be search engines, given that this is the main way to gain access to relevant sites. Google explicitly says that it rejects censorship and refuses to exclude certain sites, unless their owners desire it. (http://www.google.com/remove.html) Obviously, the standard exclusion of web crawlers via a "robots.txt" file only works if the owner of the site *wants* that site not to be covered by search engines (and if the search engine obeys that wish). Contrary to Google's claim, research seems to show that the German and French versions of Google do not list certain sites that are illegal in those countries, while google.com does (Zittrain and Edel 2002).[12]

The standard exclusions mentioned above, especially those concerning owned information, are provisions of standard law that pre-exist the Internet and indeed computer-information technology. These technical developments have caused a number of new legal issues, such as what constitutes 'copying', what kinds of personal exchange of copyrighted information is permitted (particularly urgent in the case of music, see the 'napster.com', 'mp3board.com' cases and similar organisations) and whether facilitating access to illegal information is itself illegal (e.g. linking to it)[13]. A German court has recently re-affirmed the view that Internet access providers have the obligation to block access to illegal sites - in this case sites supporting Nazi politics (Arnsberg 2004). Other courts have done the same for 'dangerous information' as we will see below.

### 3. Poisons, Viruses and Bombs: Information for the Select Few

#### 3.1. Technique Propagation
The use of information technology for malicious or plain criminal purposes is not the interest of our paper, as noted. Accordingly, the making of bombs and poisons is not our concern and neither is the writing and distribution of viruses, worms and Trojan horses. But what about information *about* such techniques? What about information about techniques that are not dependent on information technology?

---

[12] Interestingly, Google does not remove this page from its search results to suppress the news. Search engines potentially have enormous censorship power.
[13] The most useful compilation is Forum des droits sur l'internet (2003), see also Links and law (2003).

There is a threat specific to information technology that Bruce Schneier calls "technique propagation" (2000, 21f): It is not a great problem if some specialists learn how, say, to cheat coin-operated phone booths, but if someone wrote a program that is downloadable to all Java-enabled mobile phones and allows users to call for free, a significant problem would arise (or, in Schneier's example, learn how to counterfeit electronic cash). If you can download programs to hack sites or to construct viruses, then the hurdle for these kinds of things are lowered very significantly. Another practical examples of such information (that is not a technique but could be used for one) is the discovery of a security flaw in a piece of software. If one publishes this, hackers will use it. – This is the kind of problem we are interested in: should this information be spread?

This issue concerns computing techniques as well as other techniques, where the use would be disastrous. Snippets on the news like 'There are now thought to be more than 100 terrorist organisations capable of developing a rudimentary atomic bomb.' (*Guardian Weekly* 17.-23.05.2001, Vol. 164, No. 21, p. 5) are thought to be alarming (cf. also DeGeorge 2003b, 184). The same counts for the making of other bombs, for fatal poisons, biological warfare, aspects of biotechnology and nanotechnology. Dietz (1988), for example reviews print publications that advocate the use of product tampering and other poisoning methods for various purposes (revenge, murder, extortion).

Information about techniques that can cause fatal harm to large numbers of people - lately called 'weapons of mass destruction' by some - is such that we want to limit the number of persons with access to such information as much as possible. As some have argued (Laqueur 2004), the main threat of terrorism today is just in the technological means available that are so much more deadly than the anarchist bombings of latter day[14].

### 3.2. Useful Use

Can we establish that this kind of information belongs to our category of information that should never be used? I do not think that this can be declared because we can think of extreme cases where such information might become a useful tool

---

[14] This is very contentious. Most people called 'terrorists' by their opponents (and called 'freedom fighters' by their supporters) do not try to inflict maximum damage. They have qualms about innocent civilians and, particularly worries about their public image -

for a higher good. Bombs and poisons are destructive but few would say that there are no cases at all where we should use them. (If the only way to remove Hitler from power would be to poison him, thus saving many innocent lives, would there not be a moral obligation to use the poison?) Schneier says 'I am willing to live with tools that have good and bad uses, but I don't like tools that only have bad uses' (2000, 343) That sounds plausible, but it requires little imagination to find good use for almost anything.

Still, given the rarity of cases where the use of this information is justified, we must *restrict* the spread of such information and especially the spread of such techniques. An actual example is information on how to sabotage train lines that was published in the now outlawed German periodical 'Radikal'. In 2002, the national German railway operator 'Die Bahn' actively tried to prevent the spread of information from two articles of this periodical on the Internet. Its lawyers asked the search engines Google, AltaVista and Yahoo to remove any links to these articles and the companies complied. Furthermore, Die Bahn asked the Dutch Internet providers XS4ALL to stop hosting the material and the provider Indymedia to remove links to mirror sites of 'Radikal'. When the providers refused, Die Bahn took them to court in the Netherlands and won both cases (see Links and law 2003).

This problem is particularly urgent in information technology. It is not only theoretically very difficult to build an nuclear weapon, it also requires high-tech equipment and raw materials. These practical issues, not the lack of information has kept 'proliferation' of nuclear weapons at a low level. If nuclear weapons could be downloaded and copied like cracking software (and with the protection of relative anonymity), we would already have seen the first terrorist group using such a bomb.

So, if there is a piece of information that will have nasty consequences in the wrong hands, then we should make sure it remains in the right hands, if any.

### 3.3. The Select Few
Not only do we lack certainty that such information can never be used for a good purpose, we actually need to keep this knowledge around in certain circles of experts. What would we do without doctors specialising on poisons or technicians

---

the penultimate aim of any terrorist. Many factors can lead to a reduction in this humanitarian worries, not least the brutality used by either side.

specialising on bombs? (Hollywood certainly would be in trouble without the latter.) Even if we were to limit the spread of these nasty techniques, we can never be sure that they will never be used, so we need a few specialists that know as much about these techniques as possible - and we must control their use of these techniques. This is the purpose of an article for the forensic specialist like Dietz (1988). In computer terms, we need security experts, virus experts, etc. - and we know that these would be the most dangerous hackers.

### 4.   Conclusion: Ignorance is Preferable to Knowledge

At this point it seems clear that the typical demonstrations that 'information must be free' or 'Knowledge is the heritage and the property of humanity and is thus free.' (WSIS 2003a) are false. Furthermore, we need to prevent not only the spread of information that violates the rights of particular persons, but also of what I called 'dangerous information'. When this problem first became urgent, after the invention of the atomic bomb, there were already two views among the prominent physicists: Edward Teller then said 'There is no case where ignorance should be preferred to knowledge – especially if the knowledge is terrible.' (quoted in Shattuck 1996, 177). I would side with Robert Oppenheimer who thought that we should limit the spread of information about nuclear weapons as much as we can, even if we cannot put the genie back into the bottle.